# CONSIDERATIONS FOR USING REPRODUCTION DATA IN TOXICOKINETIC-TOXICODYNAMIC MODELLING

Tjalling Jager[1*], Marie Trijau[2], Neil Sherborne[3], Benoit Goussen[2], Roman Ashauer[4,5]

[1] DEBtox Research, Stevensweert, The Netherlands
[2] Ibacon GmbH, Roßdorf, Germany
[3] Syngenta, Jealott's Hill International Research Centre, Bracknell, Berkshire, RG42 6EY, United Kingdom
[4] Syngenta Crop Protection AG, 4058, Basel, Switzerland
[5] Department of Environment and Geography, University of York, Wentworth Way, Heslington, York, YO10 5NG, UK

**ABSTRACT**

Toxicokinetic-toxicodynamic (TKTD) modelling is essential to make sense of the time dependence of toxic effects, and to interpret and predict consequences of time-varying exposure. These advantages have been recognised in the regulatory arena, especially for environmental risk assessment (ERA) of pesticides, where time-varying exposure is the norm. We critically evaluate the link between the modelled variables in TKTD models and the observations from laboratory ecotoxicity tests. For the endpoint reproduction, this link is far from trivial. The relevant TKTD models for sub-lethal effects are based on Dynamic-Energy Budget (DEB) theory, which specifies a continuous investment flux into reproduction. In contrast, experimental tests score egg or offspring release by the mother. The link between model and data is particularly troublesome when a species reproduces in discrete clutches, and even more so when eggs are incubated in the mother's brood pouch (and release of neonates is scored in the test). This situation is quite common among aquatic invertebrates (e.g., cladocerans, amphipods, mysids), including many popular test species. We discuss these and other issues with reproduction data, reflect on their potential impact on DEB-TKTD analysis, and provide preliminary recommendations to correct them. Both modellers and users of model results need to be aware of these complications, as ignoring them could easily lead to unnecessary failure of DEB-TKTD models during calibration, or when validating them against independent data for other exposure scenarios.

## INTRODUCTION

Toxicokinetic-toxicodynamic (TKTD) modelling offers many advantages over descriptive methods for data analysis and the prediction of ecotoxicological effects. In fact, it is the only approach to make sense of the time dependence of toxic effects, interpret and predict consequences of time-varying exposure, and to meaningfully compare chemicals and species (Jager et al. 2006; Ashauer and Escher 2010). These advantages have been recognised in the regulatory arena, which has led to a Scientific Opinion from the European Food Safety Authority (EFSA) regarding the use of TKTD modelling for application in risk assessment of pesticides in Europe (EFSA 2018). This opinion provides a framework for TKTD modelling within this specific context. For the analysis of sub-lethal effects, the relevant TKTD models are based on Dynamic Energy Budget (DEB) theory (Kooijman 2001). There is no single unique DEB model but rather a family of closely related models, generally referred to as DEBtox, or more recently as DEB-TKTD (Jager et al. 2014; Sherborne et al. 2020). Even though these models were judged to be "not yet ready for use in aquatic risk assessment for pesticides" (EFSA 2018), their potential for supporting risk assessment was recognised. The main reason for concluding that DEB-TKTD models were not ready was a lack of published case studies for pesticides, with aquatic organisms, and including time-varying exposure. What is also lacking, yet has been overlooked, is detailed guidance on how to perform a meaningful analysis with DEB-TKTD models, considering the nature of the available toxicity data. Clearly, standard protocols for experimental tests have not been developed with the possibilities and requirements of TKTD models in mind.

In this contribution, we discuss various aspects of the link between the modelled processes in DEB-TKTD models and routine observations on reproduction. The main issues are related to the fact that the model specifies the investment into reproduction, as a continuous flux of mass or energy (see Figure 1). Experimental tests cannot directly quantify this investment and instead score egg or offspring release from the mother. For species that produce relatively small eggs, one at a time, we can usually safely ignore these details: the differences with a continuous mass flux will be rather trivial. However, many aquatic invertebrates produce clutches of eggs, and many also incubate the eggs in a brood pouch until hatching. Species orders that do both are cladocerans, amphipods and mysids, which include popular test species such as *Daphnia magna*, *Ceriodaphnia dubia*, *Americamysis bahia*, and *Hyalella azteca*. The water flea *D. magna* will serve as an example throughout this discussion. Under standard test conditions, this species produces a clutch of eggs every three days, closely linked to the moult cycle. Allocation of resources towards reproduction is continuous, so these resources are stored by the mother in a reproduction buffer, which is converted into eggs at spawning events (see Tessier and Goulden 1982). The eggs are incubated in a brood pouch and released at the next moult when a new clutch of eggs is deposited in the brood pouch. The reproduction buffer and the brood pouch incubation imply a considerable delay between the investment into reproduction and the observations on neonate release. Since this buffer and brood pouch are not represented in standard DEB-TKTD models, a temporal mismatch occurs that can easily lead to failure of TKTD modelling in risk assessments following EFSA's workflow (EFSA 2018).

These complications were recognised in the earliest DEBtox work on *D. magna* (Kooijman and Bedaux 1996), but it was proposed to largely ignore them. The modelled mass flux is instantaneously translated into a continuous egg-production rate, and that rate is cumulated over time. This cumulated egg production is then compared to the cumulated counts of released neonates over time. This procedure was subsequently used in most DEB-based data analyses. However, with the increasing interest in DEB-TKTD for pesticide risk assessment, it is important to scrutinise this procedure and explore alternatives. For pesticides, time-varying exposure is the norm, and pulsed exposure is proposed for TKTD model calibration or validation (EFSA 2018). Toxicokinetics and toxicodynamics explicitly consider the time-course of the processes underlying toxicity, so it is important to carefully match the timing of the toxic effects to the timing of the exposure events. In this contribution, we discuss these and some other issues with reproduction data, reflect on their potential impact on TKTD analysis with DEB models, and provide some preliminary recommendations.

**MAIN ISSUES IN MORE DETAIL**

We here discuss the problems for clutch-wise spawning and brood-pouch delays in a general manner. In the supporting information we present model fits on an artificial data set, providing an example for the extent of bias caused by ignoring these issues in a specific case.

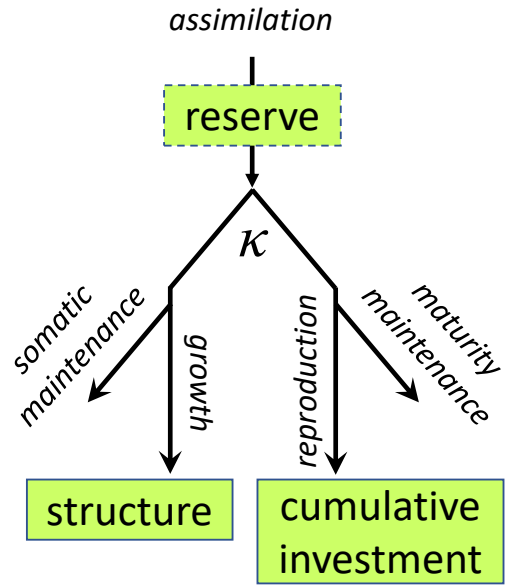

**Figure 1.** Schematic representation of the mass flows for an adult individual in a typical DEB-TKTD model. The $\kappa$ represents a constant split between the somatic pathway (growth and somatic maintenance) and the reproductive pathway (maturity maintenance and reproduction). Arrows are mass fluxes and boxes are state variables. Simplified DEB-TKTD models exclude the reserve state.

*Clutch-wise spawning*

Here, we start by considering a test species for which produced eggs are counted as reproduction; considerations for the brood pouch will be added in the next section. The basic structure of a DEB-TKTD model for an adult animal is shown in Figure 1. For any mechanistic model, the modelled fluxes and state variables may not be things that are easily measurable, requiring auxiliary hypotheses to link modelled variables to measured one (see Kooijman 2018). Egg release (spawning) is easy to score in many animals, but egg counts cannot be directly related to the reproduction mass flux in the model. The most common set of auxiliary hypotheses for DEB-TKTD modelling has always been to assume that the mass allocated towards reproduction is instantly converted into eggs (with a certain efficiency), that eggs have a constant mass, and that we can safely ignore the discrete nature of egg counts. Expressing both the modelled egg production rate and the measured egg counts cumulatively over time allows for straightforward comparison between the continuous investment flux of the model and the discrete egg counts at regular time points.

Clutch-wise spawning will lead to a step-like pattern in the data due to the presence of zero-reproduction observations at observation times between clutches (Figure 2A). The blue dots indicate observations on egg production that have been cumulated over time. If we fit a DEB model to these data, assuming continuous reproduction (i.e., no reproduction buffer in the model), the model curve (solid line) will be a compromise between all data points. However, not all data points carry information about cumulative investment in reproduction. At an observation point where an egg clutch was produced, we have unbiassed information about the cumulative investment in reproduction up to that time point. When no eggs are observed, the data point for cumulative reproduction stays the same as the one at the previous time point. There will have been investment into reproduction (stored in a reproduction buffer), but we cannot quantify it since no eggs were produced. Taking the observed reproductive output at face value, and plotting it in a stepwise graph (Figure 2A, blue dots), assumes that there has been no investment at all at these time points. The fit of a model to this stepwise data, as shown in Figure 2A, leads to bias in the parameter estimation: most data points underestimate the cumulative reproductive investment, and as a consequence, the fitted model curve is biased and lags behind the real investment into reproduction. Additionally, the residual variance will be exaggerated (compare residuals in Figure 2A and 2B), with consequences for statistical inference.

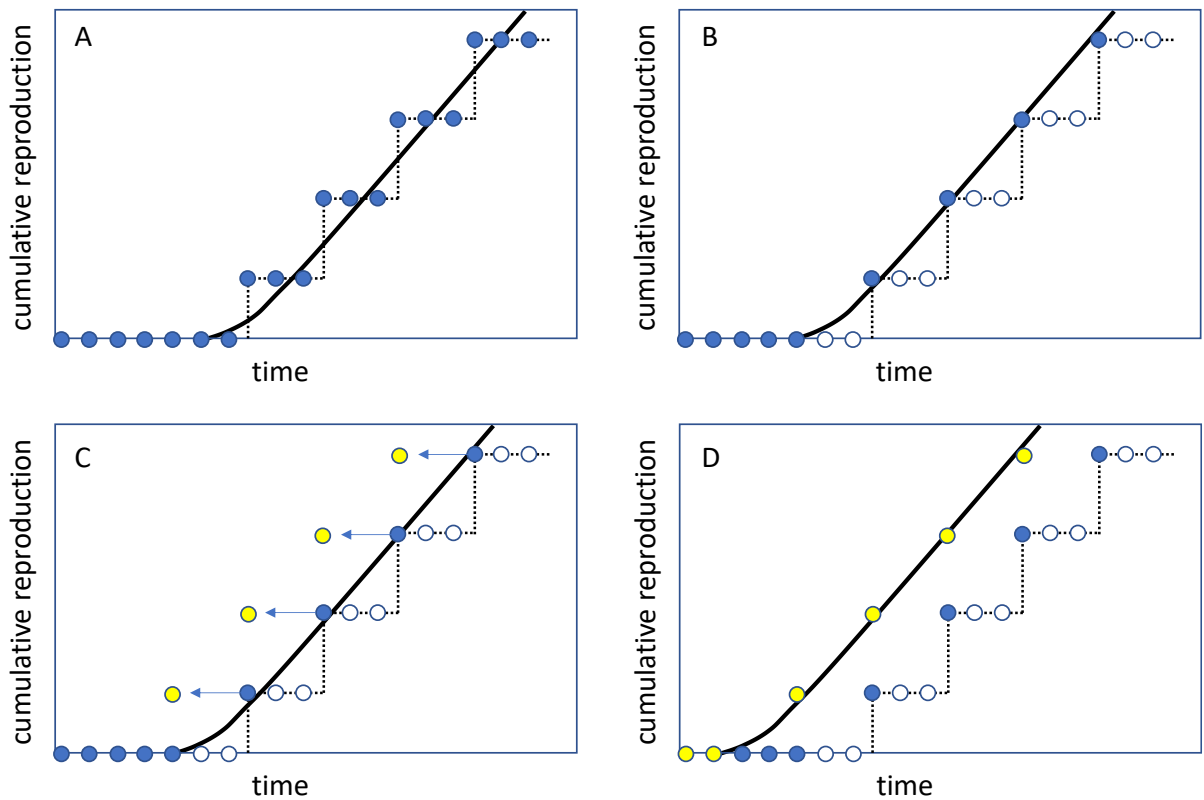

**Figure 2.** Illustrating the problems with fitting DEB models on cumulated offspring for a species that reproduce in clutches. The data points can represent an individual mother or a cohort that is perfectly synchronised. Panel A and B deal with a species for which egg production is followed over time; we can fit all data (panel A) or exclude observations without spawning events (panel B). Panel C and D deal with a species that deposits eggs in a brood pouch, and neonate release is scored. Egg production precedes neonate release by some amount of time (panel C), and it makes sense to fit the model on the (estimated) egg-production observations rather than on neonate release (panel D). Full blue circles: observed data included in model fit (release of neonates in panels C-D); empty blue circles: excluded data (no spawning or release of neonates); yellow circles: estimated production of egg clutch; black line: fitted model.

These issues can be addressed by modifying the model by including a reproduction buffer and spawning rules (see e.g., Pecquerie et al. 2009), or by censoring the data set. The latter option is the simplest and implies that we remove the time points without observed egg production (Figure 2B). This will shift the fitted curve to the left compared to Figure 2A, since the data are now more representative for the investment into reproduction. Some zero observations do carry information and need to be kept in the data set: the time points where we can be quite certain that there is indeed no investment in egg production yet since the animals are still juvenile. Deciding which zeros to keep is somewhat arbitrary but should be based on organism physiology. In general, we would propose to keep as zero all observations before the time of the first spawning event, minus the average (or initial) spacing between spawning events. The model can then be fitted to only the remaining data points (black line through filled blue dots in Figure 2B). There may also be true zeros later in the life cycle (after one or more spawning events). Especially when there is considerable toxic stress on the organisms, investment in reproduction may truly stop. This is a trickier problem since we cannot know the cumulated investment into reproduction after the last spawning event (unless we can measure the reproduction buffer directly). In line with the initial zeros for juveniles, we may also keep some zero observations after the last spawning event, accounting for the average spacing between spawning events.

This procedure of data censoring is only possible if animals were kept individually in the test, or if they are perfectly synchronised. The mean cumulative reproduction of a group of animals may not show such a clear stepwise pattern, precluding removal of zero observations. The bias in the model fit will still be there, though it will be less obvious because the means can give the appearance of continuous reproduction. Extending the model is still an option, though its benefits need to be weighed against the disadvantages of increased model complexity and the numerical issues of fitting a model with discontinuities on more-or-less continuous means.

Interestingly, the fact that not all observations on egg production carry the same amount of information also implies that the temporal resolution of reproduction data is limited by the spawning cycle. Increasing the number of observations in time will provide higher precision on the timing of the spawning events, but will not increase the number of relevant data points. Referring to Figure 2B, we will obtain more empty blue symbols, but not more of the relevant filled blue symbols.

*Brood pouch incubation*

For species that incubate their eggs in a brood pouch, we have an additional problem: what is observed in the experimental test is the number of released neonates. However, neonate release obviously occurs later in time than the production of the egg. This is illustrated in Figure 1C; the yellow points indicate the time of the production of an egg clutch, which precedes the release of neonates (filled blue points). For TKTD modelling, it makes sense to fit the model to the (estimated or observed) egg production events, excluding the zeros as explained in the previous section. These data points will have a much closer correspondence to what is modelled, namely the resource investment into reproduction. Shifting the time vector of the reproduction data was (to our knowledge) first proposed for *D. magna* by Jager and Zimmer (2012). These authors shifted the model predictions (rather than the data, as illustrated in Figure 1D) by the average length of the intermoult period.

The brood-pouch delay can be incorporated by shifting the model prediction (such that the model output represents neonate release) or by shifting the data (such that the observations represent egg production). Shifting the model prediction is only attractive when the shift (i.e., the time between egg production and neonate release) can be taken constant over the test, across individuals, and across the treatments. If the shift is not constant, it is more transparent to modify the time vector of the reproduction data to represent the most-plausible points of egg production. In *D. magna*, neonate release and egg production occur at the moults, which offers a straightforward approach to deal with the data: we can shift the observations on neonate release back to the previous release event (or the previous moult), for each individual separately. Again, such a procedure of data censoring would only be possible if animals were kept individually in the test, or if they are perfectly synchronised.

## RELEVANCE FOR RISK ASSESSMENT

In previous studies, it was shown that DEB-TKTD models can provide fully acceptable fits on combined body size and reproduction data by ignoring the complexities of clutch-wise spawning and brood pouch incubation. However, almost all published DEB-TKTD analyses to date have been conducted for constant exposure. The parameter estimates will have been biased to some extent, but that bias has a limited effect on the model fit. For pulsed exposure, however, there is a more pressing need to get the timing aspects right. This is easiest illustrated for the brood-pouch delay, as demonstrated in Figure 3. In this case, we assume that the chemical has rapid damage dynamics: effects respond quickly to changes in exposure concentration. Furthermore, the exposure pulse is assumed to be stressful enough to completely stop investment in reproduction, but only during the pulse. In this example, the pulse hits the animal just after the first brood is released. The investment in reproduction stops, but the eggs for the second brood are already in the brood pouch; they were produced before the pulse. Therefore, even though the pulse has immediately and completely stopped the investment in reproduction, the second brood will show no effects of chemical stress. Only in the third brood does the effect show up. However, unlike the complete stop of investment into reproduction, the observed effect at the third brood will not be complete, since there was some investment into reproduction just before and after the pulse. If we would fit a DEB-TKTD model to such an effect pattern (the filled blue symbols in Figure 3), without considering the delay caused by the brood pouch, we would incorrectly conclude that this chemical has rather slow damage dynamics (delayed effects) since toxic effects only show up several days after the pulse exposure.

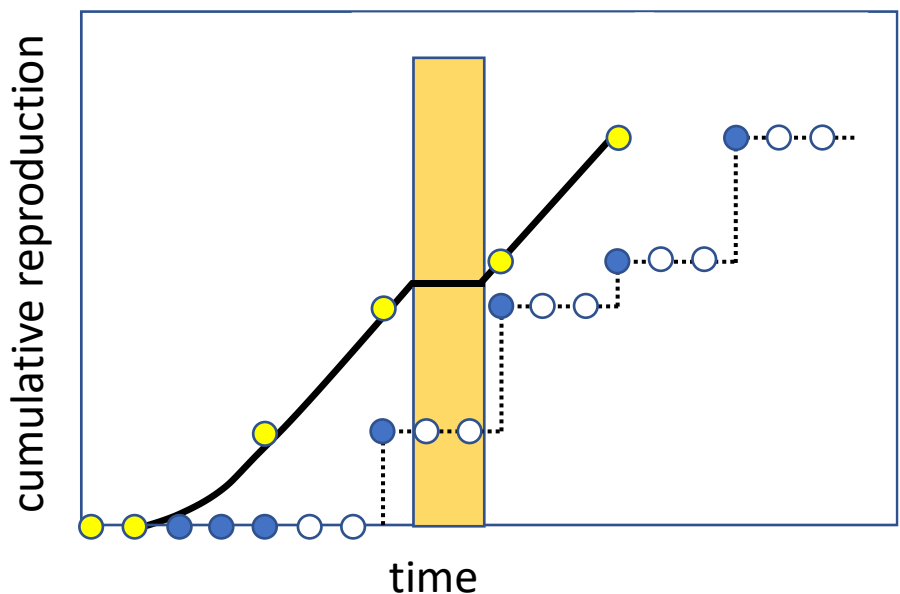

**Figure 3.** Pulse exposure for a chemical with rapid damage dynamics that completely stops reproduction during the exposure event. The bar indicates the timing of the exposure pulse. Filled blue points are observed released neonates, yellow points are estimates for the egg-production events, and the solid line is the modelled investment into reproduction.

To what extent will this influence risk assessment applications of the model? The answer will be highly dependent on the species and toxicant, but also on the type of extrapolations that are needed for a specific risk assessment application. Previous work with a TKTD model for survival showed that the sensitivity of model parameters depends on the exposure pattern used for extrapolation (Ashauer et al. 2013). Similarly, we expect that bias in model parameters for DEB-TKTD models will affect some predictions more than others, depending on the exposure pattern. As is true for all extrapolations, uncertainty will increase the more the exposure conditions in the calibration data set(s) differ from the scenario for model prediction. It is good to realise, that EFSA's scientific opinion (EFSA 2018) provides a number of safeguards against biased model predictions. Indeed, the DEB-TKTD model needs to be calibrated on experimental data for the specific species-chemical combination, and subsequently validated on a data set with a different exposure profile to prove that the model is able to extrapolate across exposure conditions. There are quality requirements on the data sets, and on the goodness-of-fit in the calibration and validation stages. If the data are of sufficient quality, and if the model provides a good correspondence to the data sets in both the calibration and validation stages, we are confident that it will also produce meaningful predictions for untested exposure scenarios. In any case, we are certain that it will provide a more biologically-relevant risk assessment than traditional descriptive procedures without DEB-TKTD models.

A matter of greater concern is that ignoring the complexities for reproduction data will likely lead to rejection of many model analyses for risk assessment. We will usually fit multiple endpoints from the same toxicity test simultaneously: reproduction, body size and survival. Under pulsed exposure, ignoring the complexities of clutch-wise spawning and brood pouch delays can easily produce artificial inconsistencies between the effects on the various endpoints, resulting in poor overall model fits in calibration, and poor predictions in validation. It is important to critically evaluate TKTD models, based on their performance in calibration and validation on different data sets. However, it is also important to realise that rejection of TKTD models implies falling back to descriptive methods (e.g., static dose-response curves and time-weighted average exposure), which are not held to the same high standards (Jager and Ashauer 2018). The problem here is that the TKTD model could be rejected for the wrong reason: it would not be a failure of the model itself, but rather an oversimplified link between the modelled processes and the nature of the observations (the auxiliary hypotheses).

As explained, the issues regarding clutch-wise spawning and brood-pouch delay may be addressed by censoring and shifting the data set, or by extending the model. Modifying the data set is often the simplest solution but may cause concern in a field where observations are often viewed as objective and 'true'. Therefore, it is important to stress that such data modifications are not intended to fix a poor model fit, but rather to objectively make the data reflect the modelled properties (they result from essential auxiliary hypotheses). Not observing an event does not provide information on the mechanisms underlying the observations at the event. Such censoring would be specific for a species and a test protocol, but would not depend on the chemical, possibly with some exceptions (one is treated in the next section: when a chemical is taken up by the eggs in the brood pouch and affects development). Keeping the data as is, and extending the DEB-TKTD models, is also a possibility. Including a brood-pouch delay into the model is simple, as long as the required shift is constant. Including a reproduction buffer in the model is also feasible (for an example, see Pecquerie et al. 2009), requiring rules for spawning decisions. However, modelling a buffer implies more complex model code, numerical issues due to switches in the model, and likely additional model parameters.

**ADDITIONAL ISSUES**

Beyond the clutch-wise spawning and brood-pouch delay there are several additional complications with reproduction data that are important to consider, but which will not be discussed here in the same level of detail.

*Non-constant cost per egg*

Simplified DEB-TKTD models usually rely on the assumption that, within a species, all eggs are equal. In other words: the cost for a single egg, in terms of energy or mass, is constant. Therefore, we can easily use observations on egg numbers as a proxy for the investment into reproduction. However, there are cases where this assumption is violated. For *D. magna*, for example, the investment per neonate depends on the mother's age or size, and on her feeding status (Gabsi et al. 2014). This will cause some bias in the DEB model parameters, but more important for risk assessment is that toxicant stress may also affect offspring size. For example, we may expect that chemicals affecting assimilation would have the same effect as food limitation (increasing offspring size). Furthermore, some toxicants have been shown to specifically decrease offspring size, potentially biasing interpretation of the toxicant's risks (Hammers-Wirtz and Ratte 2000). It would not be too hard to modify the model to account for changes in the cost per egg. However, parameterisation of such an extension requires information about egg or neonate size throughout the toxicity test. Such measurements are not foreseen in standard test protocols and may substantially increase testing efforts.

*Dead neonates and aborted eggs*

For animals that deposit their eggs into a brood pouch, we may observe egg abortion or dead neonates, especially under severe toxicant stress. The regulatory endpoint of the Daphnia reproduction test is the number of live offspring at the end of the test (OECD 2012). DEB models are concerned with the allocation of resources, and ignoring the investment represented by aborted eggs and dead offspring has the potential to bias bioenergetic analyses. Abortions and dead offspring can already occur in the controls of a toxicity test, especially in the first brood. Counting only the live offspring would bias the estimation of the DEB model parameters, since the investment into reproduction will be underestimated. However, the abortions and deaths may also be an effect of the chemical; a situation that requires more careful consideration. In principle, there may be two mechanisms by which the toxicant can affect developing eggs, leading to abortions or dead neonates. The first route is through the mother: the mother is affected by the chemical stress, which leads her to produce eggs that are compromised in a way that hampers their development. The second route is through the egg: the egg is exposed in the brood pouch, either through maternal transfer of toxicants, or by uptake from the medium in the pouch, which affects its development.

If the effect is on the mother, and if there is only a direct effect on reproduction, no special attention is required. If only live offspring are counted, there is no relevant difference between reduced survival of embryos and reduced resource investment into reproduction, and the same DEB-TKTD model formulation can be used for both cases. The situation is, however, different when there is also an effect on growth. Mechanisms of action which affect growth will, in the DEB-TKTD context, always have consequences for the investment in reproduction. In such cases, we need to separate the effects on reproduction into a reduction in investment (to which all offspring, alive or dead, contribute) and a reduction in survival of the embryos. Failure to do so can easily lead to a mismatch between the effects on growth and reproduction.

If the effects are caused by exposure of the eggs in the brood pouch, the situation is more complex. This case requires us to consider the egg as a dynamic system with its own TKTD model. Furthermore, the brood pouch delay becomes irrelevant; in the exposure scenario of Figure 3, we would expect to see an effect on the second brood already. The timing of the observed effects, relative to the exposure pulses, may thus reveal which route is most likely.

*Deaths of the mothers*

The standard test protocol for Daphnia reproduction (OECD 2012) states that, for accidental deaths (not concentration dependent), the replicate with the dead mother should be excluded from the analysis. If the deaths are concentration dependent, it is an effect of the compound, and the replicate should be left in; there will be data points with zero neonates for this replicate after the mother has died. For DEB-based modelling, both options do not make sense: mothers that do not survive till the end of the test still contain information on the reproduction rate, and we should not add information to the data set that has no basis in observation, i.e., assuming an explicit zero reproduction rate for dead mothers. The same concerns were also raised for classical dose-response modelling (Delignette-Muller et al. 2014). All mothers contribute information on reproduction up to the point where they die, and no further; the subsequent reproduction observations are 'missing data points' and not zeros. Since DEB-TKTD models are fitted on observations over time, premature deaths are not problematic for the analysis, as long as we correctly weigh in the number of contributing mothers in model calibration.

Removal of replicates will not bias a model analysis; it is only inefficient as valuable information is discarded. Including zeros after the mother has died is more problematic as it has the potential to completely disrupt the calibration and validation procedures. One problem is that effects on reproduction can no longer be interpreted as (only) an effect on the energy budget, and hence effects on reproduction cannot be matched to effects on growth anymore. A second problem occurs under pulsed exposure, since effects on reproduction tend to be reversible, but effects on survival are not. Keeping reproduction of dead mothers into the data set will thus provide a biased view on the potential for recovery. Clearly, deaths of mothers is a relevant test endpoint, but this needs to be modelled separately, using a survival module within the DEB-TKTD model, and not by modifying the reproduction data.

*Statistical issues*

Fitting models requires a model for the process (here a DEB-TKTD model) but also a model for the residuals, describing the difference between model curve and observations. Selecting an appropriate statistical model for reproduction data is far from trivial (Jager and Zimmer 2012). In many cases, the observations are made on the same cohort of animals, followed over time. Therefore, the observations are not independent, which is deteriorated further by cumulating reproduction over time. Reproduction observations are counts of eggs or neonates and hence discrete. The residual variance usually increases with the mean reproduction, but even worse: the residuals are not the result of random measurement error but due to biological variation and the fact that the DEB-TKTD model is a simplification of reality.

Due to this list of problems, there is currently no appropriate statistical model for reproduction data over time. Several authors have proposed a negative binomial distribution for cumulative reproduction counts (Delignette-Muller et al. 2014), also specifically for DEB-TKTD modelling (Billoir et al. 2011). While this indeed addresses several of the issues above (discrete data and residual variance increasing with the mean), it does not solve the more important issues (dependence and the nature of the error), while requiring an additional parameter to be estimated from the data. For now, we therefore propose to stick to the familiar likelihood function based on the normal distribution for independent observations. To account for the increase of the residual variance with the mean, we propose a mild transformation such as square-root transformation. This precludes the fit to be dominated by the high values for the cumulative reproduction. Log-transformation also does this but is more problematic due to the initial zeros, and because it places large emphasis on the appearance of the first brood. Because of the limited temporal resolution in most tests, and the possibility for the first brood to deviate (e.g., smaller individuals as in *D. magna*), this could lead to unrealistic fits.

Clearly, this statistical model is a poor representation the error structure in the data; more work is needed to develop better matching alternatives (without complicating the model). We do not expect that these limitations will lead to bias in the model fits or in the model predictions. However, the confidence intervals on the model parameters and the model predictions will be compromised, which should therefore be interpreted with care.

**RECOMMENDATIONS**

The link between model and observations requires closer scrutiny for DEB-TKTD models, especially for the endpoint reproduction. Most test designs currently follow the prescriptions in standard test protocols, which were never intended for mechanistic model analysis. The EFSA opinion on TKTD models (EFSA 2018) provides no guidance on this mismatch, which would also have been complicated since one single strategy is unlikely to fit all species, all stressors, and all test designs. In Table 1, we provide some preliminary recommendations on the various issues we put forward. Proper guidance would need to be tailored to the peculiarities of the test species, and the possibilities for experimental testing. For example, for some species it is not practically feasible to follow individuals over time. Furthermore, some recommendations involve modifications of the test design; developing proper guidance would be most efficient in conjunction with a revision of the standard test protocols.

Ignoring the issues listed in Table 1 will always cause bias in the model parameters. It is unclear to what extent this will also lead to bias in model predictions; this will likely be highly case specific. Such bias tends to go unnoticed in the control fits, and often also in fits for effects under constant exposure. However, we can be certain that there will be cases that are severe enough for the DEB-TKTD model to fail in explaining the observations for pulsed exposure. In the supporting information we offer a case study, illustrating the potential bias caused by ignoring clutch-wise spawning and brood-pouch delays. This is just one example, but it shows that especially the brood-pouch delay can provide a distorted picture of the effects due to pulsed exposure. More example studies would allow a clearer picture of the extent of this bias, and the cases that cause the greatest concern. However, uncertainty about the extent of the impact should not be used as an excuse to perpetuate an inherently flawed set of auxiliary hypotheses.

At this moment, TKTD modellers will need to decide how to use reproduction data on a case-by-case basis. We strongly advise them, however, to explicitly mention and motivate that choice in their reporting, for each of the issues in Table 1, including when no modifications are used. As TKTD models are receiving increasing attention in the risk-assessment community, more and better data are bound to become available. In due time, this will allow more structured guidance to be developed. It is, however, of paramount importance to ensure that all data sets used within an analysis are treated in the same manner. Using a censored data set for calibration and an uncensored one for validation is bound to cause problems. This warning also extends to the use of DEB parameters from the add-my-pet library, which is needed to apply the most extensive DEB-TKTD model variants (Sherborne et al. 2020).

**Table 1.** Practical recommendations for the various issues with reproduction data. The last column summarises the potential for failure of the model calibration and validation, and our subjective evaluation for the potential bias in model parameters.

| Issue | Recommendation | Note | Consequences of ignoring the issue |
|---|---|---|---|
| Clutch-wise spawning | Censor data set if animals are followed individually or are closely synchronised | As alternative, we can include the reproduction buffer into the model; this requires rules for the timing of the spawning events | Medium potential for failure or bias |
| Brood pouch delay | Shift data set or model output in time | Shifting the model output is only possible if the required shift is constant over time and across treatments; do not shift data or model when the chemical is taken up by the egg and affects development in the brood pouch | High potential for failure and bias under pulsed exposure |
| Non-constant egg costs | Follow egg/offspring size in the test | Counts on reproductive output would need to be scaled to a reference offspring size, or investment per offspring in the model must be variable | High potential for failure and bias only if egg costs are concentration dependent |
| Aborted eggs/dead neonates | Count and classify all reproductive output | Which counts to use, and how to use them, can be decided on a case-by-case basis; effects on total reproductive output are best separated from effects on egg development and survival | High potential for failure and bias only if abortions/deaths are concentration dependent; low potential when effect is through the mother and on reproduction only |
| Death of mothers | Include reproduction by all individuals, for as long as they are alive | Reproduction in an interval needs to be weighed according to the (average) number of females alive in that interval; deaths need to be covered by a survival module in the DEB-TKTD model | High potential for failure and bias only if reproduction for dead mothers is set to zero |
| Statistical issues | Use likelihood based on the normal distribution with a mild transformation | | Low, but confidence intervals need to be treated more qualitatively |

## CONCLUSIONS

TKTD models offer a powerful means to interpret and predict toxicity, accounting for the development of the individual organism over time as well as the time-dependency of exposure. This is a huge benefit for pesticide risk assessment, since it is impossible to experimentally test all potentially relevant exposure situations. However, this power comes at a price. While endpoints such as survival and body size allow for a relatively straightforward link to modelled state variables, this link requires closer scrutiny for the endpoint reproduction. This is especially true when a species releases eggs in clutches, and even more so when eggs develop inside the mother's brood pouch. The primary purpose of this paper is to raise awareness of these complications amongst modellers and users of model results, and point out the potential for bias in model analyses. In a broader sense, this paper embodies an invitation to carefully consider the need for auxiliary hypotheses in mechanistic modelling in general, as integral part of the empirical cycle (Kooijman 2018).

EFSA's scientific opinion provides ample insurance against biased model predictions because it emphasises model validation with independent experiments. However, we expect that ignoring these complications will lead to failure of DEB-TKTD analyses in many cases, in the validation stage or even already during calibration. This is unnecessary since the reasons for this failure would be an avoidable mismatch between the data collected from standard test protocols and the information needs for DEB-TKTD models. It is furthermore unhelpful, because rejection of TKTD models implies falling back to traditional descriptive methods, which do not solve these issues but rather add a series of more fundamental problems (Jager 2011). We hope that the recommendations provided in Table 1 allow for more successful application of DEB-TKTD models.

## Acknowledgements

We acknowledge funding from Syngenta. All authors have an interest in the (regulatory) acceptance of DEB-TKTD models.

## Data Availability Statement

This paper contains no data or calculation tools.